# Scaling LLM Agents for Materials Design through Hierarchical Collective Reasoning

Jaehwan Choi[1] and Yousung Jung[1,2*]

[1] Department of Chemical and Biological Engineering (BK21 four), and Institute of Chemical Processes, Seoul National University, 1 Gwanak-ro, Gwanak-gu, Seoul 08826, Korea

[2] Institute of Engineering Research, Seoul National University, 1 Gwanak-ro, Gwanak-gu, Seoul 08826, Korea

* Email: yousung.jung@snu.ac.kr

## ABSTRACT

Materials design is frequently constrained by competing requirements: improving a functional property can compromise stability, introduce defects or restrict synthesis. Although conventional generative models can produce stable, novel crystal structures, translating detailed natural-language design instructions into candidates that reconcile these constraints remains challenging. Here we introduce HiMatGen, a framework that scales large language model (LLM) agents through hierarchical collective reasoning. Built from GPT-5.6 Terra, HiMatGen connects 100 investigators organized into discussion pods across ten scientific domains with representatives connected with domain-specific computational and retrieval tools. Representatives investigate proposals, exchange evidence across domains and return unresolved questions and computational findings to their pods. This bidirectional exchange allows specialist disagreements to drive structural revisions and alternative designs throughout exploration. Across six fixed chemical systems, HiMatGen produces 1.9 times as many final crystal candidates and 1.7 times as many stable, unique and novel (SUN) structures as a tool-enabled single-agent baseline powered by the frontier model GPT-6 Astra. Property-directed tasks show that the complete HiMatGen workflow outperforms same-model single-agent and independent-generation controls, while comparisons with MatterGen and Chemeleon2 highlight its ability to generate diverse crystals satisfying joint functional and chemical requirements. Alongside these design capabilities, hierarchical discussion reduces debate-stage token usage by a factor of 22 relative to full debate in a paired comparison. Together, these results establish HiMatGen as a broadly applicable approach to materials design in which distributed scientific knowledge, computational evidence and iterative revision remain connected as the agent population grows.

## INTRODUCTION

A central challenge in materials design is to satisfy functional requirements that often compete with one another and with the chemical constraints governing stability and synthesis. In lead-free halide perovskites, for example, replacing lead to reduce toxicity must be balanced against preserving optoelectronic performance and chemical stability.[1,2] Navigating such trade-offs requires expertise from several fields to inform a shared design process, in which different scientific perspectives help evaluate an initial proposal and guide its revision.

Computational approaches to materials design have increasingly used generative models to explore crystal structures and propose candidates with desired properties.[3,4] Evaluating and refining these candidates requires interpreting design requirements alongside chemical knowledge and computational evidence. Language-model agents have emerged as a means of connecting these elements: they can interpret natural-language objectives, consult literature, operate computational tools and revise proposals in response to the resulting evidence.[5–10] Multi-agent frameworks extend this approach by distributing research across specialist roles and coordinating their activities through task decomposition and supervised workflows.[11,12] Scientific-agent frameworks further incorporate structured critique and iterative hypothesis refinement into the research process.[13–15] As the number and diversity of participating agents increase, effective coordination must preserve distinct lines of inquiry, enable specialists to challenge one another's assumptions and carry useful findings into subsequent design decisions. Scaling collective scientific reasoning therefore depends on how information is exchanged and acted upon throughout the research process.

In this context, OpenAI's recent Navier–Stokes effort illustrates the potential of coordinated exploration at scale. Groups of tool-enabled agents pursued different approaches, and useful intermediate insights were consolidated across groups into follow-up prompts that guided subsequent investigation.[16] HiMatGen was developed independently, and its architecture and reported agent trajectories were completed before the public release of that effort on 8 September 2026, while this manuscript was being finalized. The two approaches differ in coordination architecture and feedback granularity: whereas the OpenAI effort consolidated insights across parallel groups to guide subsequent exploration, HiMatGen maintains a persistent two-level hierarchy in which candidate-linked evidence and unresolved questions are repeatedly exchanged between domain representatives and their pods. This bidirectional exchange keeps coordination active as individual candidates evolve, allowing evidence obtained in one domain to guide cross-domain investigation, structural revision and alternative design.

Here we introduce HiMatGen, a materials-design framework that treats hierarchical coordination as an integral component of collective scientific reasoning. HiMatGen organizes 100 investigators into small pods across ten scientific domains to develop and debate alternative chemical hypotheses, while persistent domain representatives guide their development through domain-specific tools and cross-domain discussion. Representatives evaluate competing interpretations, select follow-up tests and return findings or unresolved questions to their pods. Investigators use this feedback to propose structural revisions and alternative designs for further evaluation, allowing higher-level decisions to shape local exploration and emerging ideas to redirect the broader research effort. With the number of discussion rounds held fixed, repeated all-to-all debate generates peer-report traffic that grows quadratically with investigator population,[17] whereas fixed-size pods keep local report sharing linear, with additional communication among representatives and feedback to investigators. HiMatGen thus sustains broad exploration and iterative refinement while limiting the repeated transmission of detailed scientific arguments.

We evaluate HiMatGen through two complementary settings: discovering new crystal structures within specified chemical systems and designing materials under competing functional requirements and chemical constraints across property-directed tasks. The resulting portfolios are compared with those from single-agent research, independent generation followed by integration and generative models, using thermodynamic qualification and structural-matching criteria. Analysis of recorded exchanges reveals how specialist objections guide alternative structures, computational tests and revised design decisions. A paired comparison with full debate, initialized from the same proposals, assesses whether hierarchical coordination reduces communication overhead while maintaining productive materials exploration. Together, these analyses assess whether hierarchical collective reasoning translates complementary scientific perspectives into diverse, computationally qualified materials portfolios.

# RESULTS AND DISCUSSION

## Hierarchical collective reasoning for materials design

HiMatGen connects broad hypothesis exploration by 100 investigators with sustained, tool-grounded research through a hierarchy that supports cross-domain discussion and bidirectional feedback. The investigators are distributed across ten scientific domains: composition and bonding; structure and symmetry; thermodynamics; synthesis and precursors; kinetics and metastability; defects and surfaces; functional properties; characterization; cross-domain transfer; and data-driven modeling (**Figure 1a**). Each domain comprises two five-agent pods whose members bring complementary subdomain and career perspectives, with one member of each pod serving as the principal investigator (PI) and pod lead (**Figure 1b**). Investigators first propose hypotheses independently and then discuss them within their pods. The two PIs subsequently engage in two rounds of within-domain discussion, drawing on contributions from both pods to examine competing proposals and objections. They identify alternatives and follow-up questions for the domain representative to investigate, with consolidation designed to preserve distinct rationales and useful controls for subsequent tool-based evaluation.

The ten persistent domain representatives form the tool-enabled research layer (**Figure 1b**), with their scientific roles and associated tools summarized in **Table S1**. Each can construct and edit crystal structures, inspect geometry, retrieve Materials Project[18] data, evaluate candidates with MACE[19] and CRISP[20] models, and use tools for synthesis assessment[21] and literature retrieval. Domain identities guide which questions receive attention, while the registered scientific-tool catalog is available across the representative layer. A structural concern can therefore be examined by a composition or functional-property representative when it becomes relevant to that representative's investigation. Candidate records link structures to tool observations and parent–child revisions, allowing subsequent decisions to draw on the history of each design.

Following this initial proposal phase, research proceeds through three research stages at the representative level, separated by two downward feedback stages (**Figure 1c**). Representatives investigate candidates using scientific tools, seek input from other domains and return specific findings or unresolved questions to their own pods. At each feedback stage, all 100 investigators receive relevant evidence and propose corrections, controls or alternative designs. The subsequent representative stage draws on these replies and the completed cross-domain exchanges to guide further investigation. This feedback cycle keeps the broader investigator population engaged as candidates evolve, allowing unexpected coordination environments or unfavorable computational results to prompt new hypotheses and structural revisions.

We used GPT-5.6 Terra as the LLM backbone throughout HiMatGen to examine whether hierarchical coordination could extend the materials-design capabilities of a model suited to repeated, large-scale agent interactions. This choice enabled sustained exploration across the investigator population while allowing the contribution of collective reasoning to be evaluated against stronger single-agent models such as GPT-5.6 Sol and GPT-6 Astra.

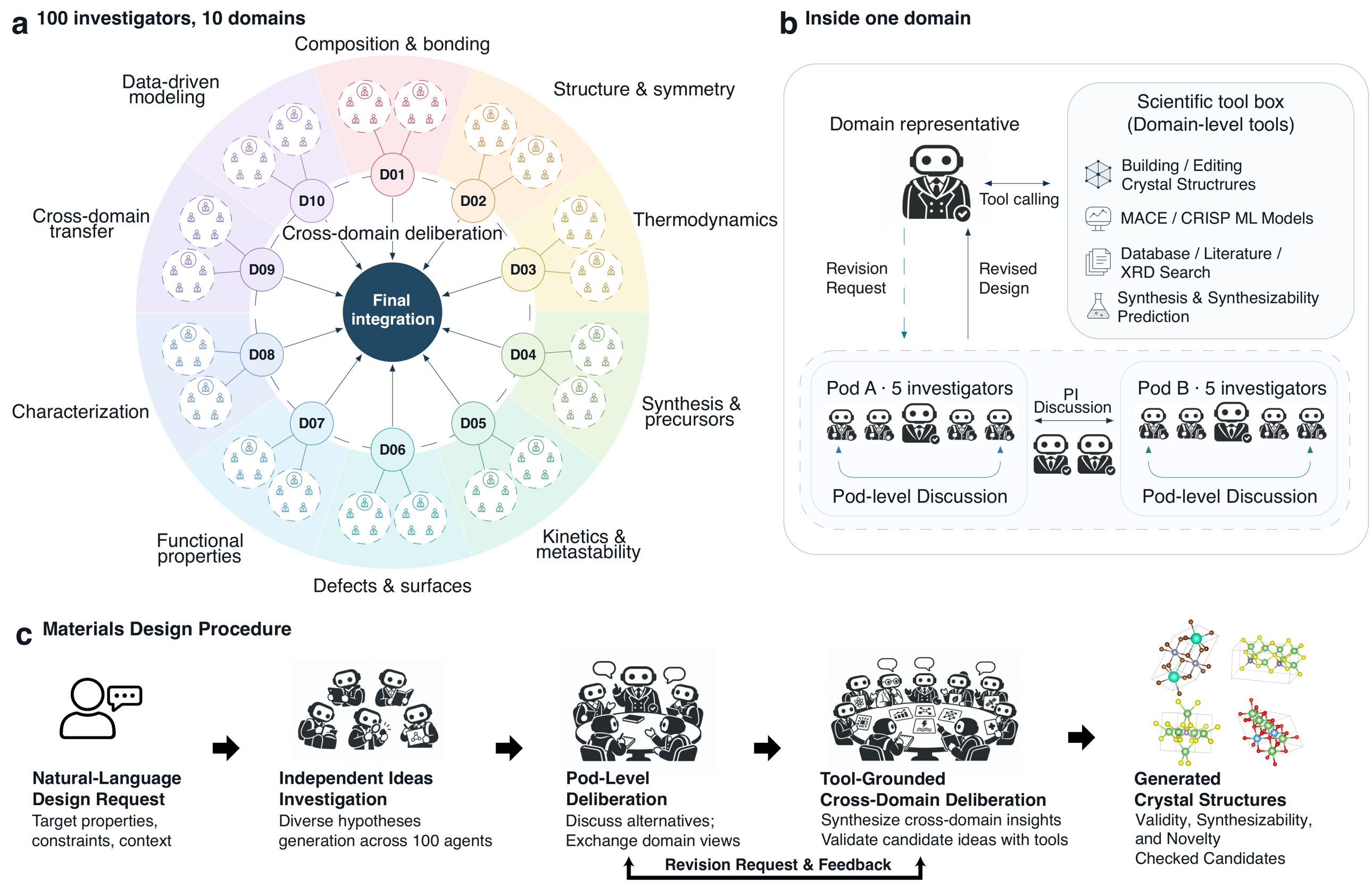

**Figure 1. Hierarchical collective reasoning in HiMatGen.** (a) One hundred investigators explore ten scientific domains, each represented by a persistent domain agent. Domain representatives exchange candidate-specific evidence before final integration. (b) Each domain contains two five-agent pods, with one member of each pod serving as the principal investigator. Following pod-level discussion, the two PIs discuss competing proposals and identify alternatives for the domain representative to investigate. The representative develops and evaluates these ideas using the scientific-tool catalog and returns findings or revision requests to the investigators. (c) A natural-language design objective initiates independent exploration, local deliberation and tool-grounded cross-domain research, with feedback supporting iterative structural revision. The process produces explicit crystal structures and associated evaluation records. The detailed stage sequence and tool permissions are provided in **Supplementary Note 1**.

## Discovering novel crystal structures within chemical systems

To assess HiMatGen's ability to expand structural exploration across diverse chemistries, we evaluated crystal discovery within six elemental systems spanning intermetallic, halide, oxide and sulfide materials: Al–Li–V, Al–Cu–Mg, Cs–Sn–Br, Li–Ti–O, Li–P–S and Li–Ge–S. Each method received known structures from the Materials Project as starting points and was allowed to explore composition changes and structural redesign within the specified elements. The search therefore encompassed new compositions as well as alternative structures at known compositions. We compared all methods across the six systems, with one research trajectory per method and system.

We assessed exploration through three successive endpoints: proposed crystal candidates, stable and unique (SU) structures, and stable, unique and novel (SUN) structures (**Figure 2**). Proposed candidates comprise final published CIFs that passed basic structural validity checks. SU counts comprise distinct structural groups containing at least one candidate with converged MACE relaxation and an energy above hull $(E_{\mathrm{hull}})$ at or below the specified threshold. SUN additionally requires structural nonmatching to both the Materials Project database and the supplied references. We used 0.10 eV per atom as the primary threshold and 0.20 eV per atom for sensitivity analysis, distinguishing proposal volume from the number of distinct, energetically qualified and reference-unmatched structures.

HiMatGen generated 220 final crystal candidates across the six chemical systems, averaging 36.67 per system,

compared with 19.00 for a tool-enabled single agent using GPT-6 Astra and 3.83 for Independent SWARM (N=100) (**Figure 2a**). Tool-enabled single agents using GPT-5.6 Terra and GPT-5.6 Sol generated averages of 3.50 and 5.33 candidates per system, respectively. The comparison with Independent SWARM is particularly informative because both approaches begin with independent proposals from 100 investigators. Independent SWARM then assigns these proposals to a single tool-enabled agent for further development and integration, whereas HiMatGen maintains ten domain representatives who investigate alternatives, exchange findings and draw on continued feedback from their pods. The larger candidate portfolio produced by HiMatGen highlights the value of sustained, distributed investigation in developing initial ideas into explicit crystal structures.

HiMatGen retained a larger portfolio after energetic screening, structural deduplication and reference matching (**Figure 2b,c**). At the primary threshold of 0.10 eV per atom, it retained averages of 8.83 SU and 8.00 SUN structures per system. The corresponding SUN averages were 2.50, 4.17 and 4.83 for tool-enabled single agents using GPT-5.6 Terra, GPT-5.6 Sol and GPT-6 Astra, respectively, and 1.67 for Independent SWARM. HiMatGen thus produced 4.8 times as many SUN structures as Independent SWARM and 1.66 times as many as the frontier GPT-6 Astra single-agent baseline. Raising the threshold to 0.20 eV per atom increased the HiMatGen SUN average to 13.17, compared with 6.67 for Astra and 4.33 for Sol. Its larger proposal portfolio translated into a substantially larger absolute number of distinct, energetically qualified structures unmatched to the reference collections under both screening thresholds.

The system-level results show how individual chemistries contributed to this overall expansion (**Figure S2**). At the primary threshold, HiMatGen retained 17 SUN structures in Al–Cu–Mg, its largest portfolio among the six systems, and ten in Li–Ti–O. The single Sol and Astra agents performed better in Cs–Sn–Br, each retaining five SUN structures compared with three for HiMatGen. These differences suggest that the value of sustained multi-domain investigation depends on the chemical setting, with particularly substantial gains in systems where HiMatGen developed a broader range of qualified alternatives. Threshold sensitivity and composition-space maps are provided in **Figures S3** and **S4**.

To visualize the structural diversity explored by HiMatGen, we jointly embedded the 700 supplied reference structures and 220 generated candidates using a species-agnostic geometric fingerprint followed by t-SNE (**Figure 2d**). HiMatGen candidates occupy multiple neighborhoods around and between reported structures, illustrating the range of geometries explored within each chemical system. Colors distinguish the six chemical systems, while pale circles and dark stars denote reported structures and HiMatGen candidates, respectively. This visualization complements the SU and SUN counts by illustrating the range of structural geometries explored by HiMatGen. Additional embedding analyses and individual-system views are provided in **Figure S5**.

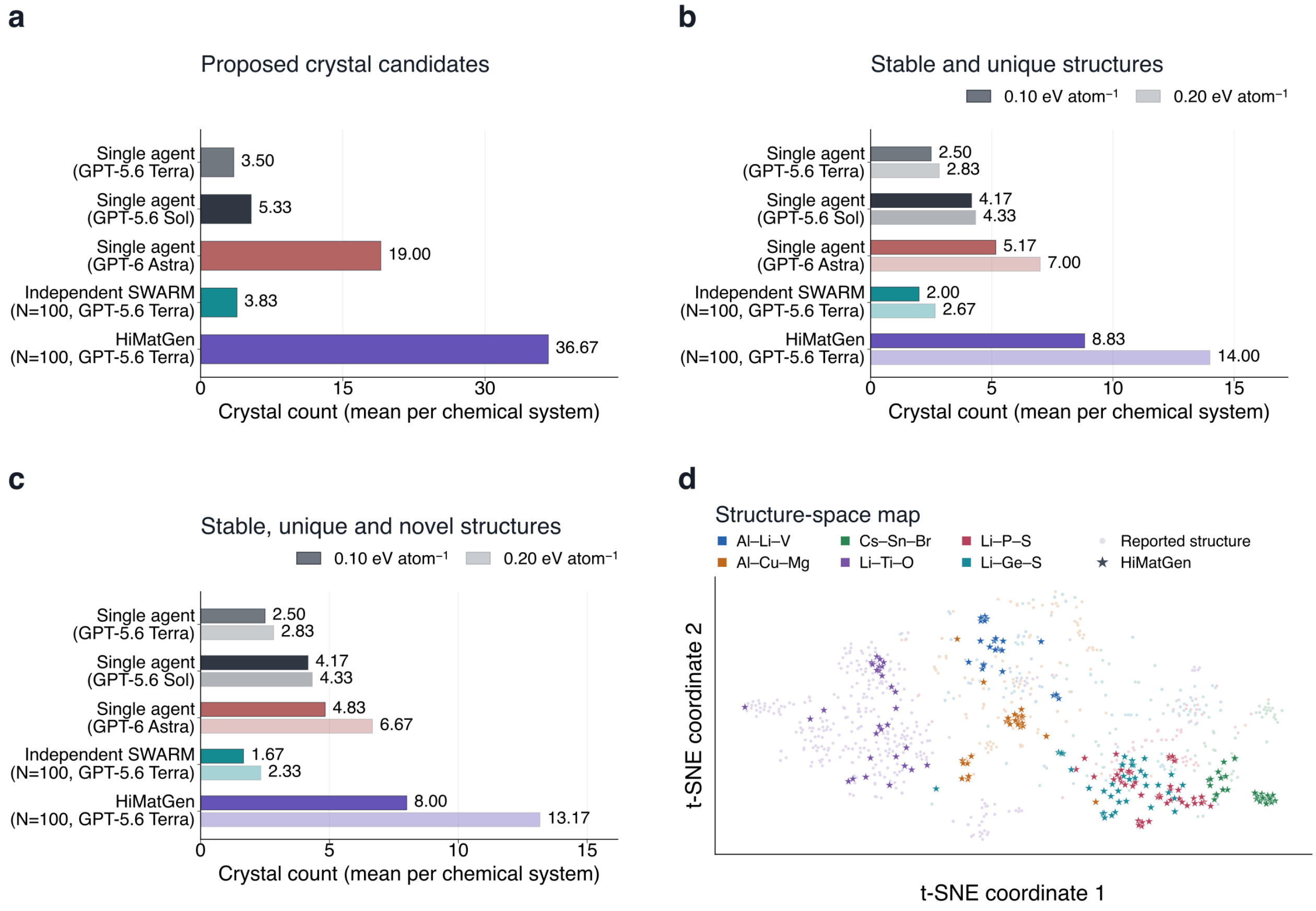

**Figure 2. Crystal discovery across six chemical systems.** (a) Mean number of proposed crystal candidates passing

basic structural validity checks. (b) Mean number of stable and unique (SU) structures and (c) stable, unique and novel (SUN) structures at energy-above-hull thresholds of 0.10 and 0.20 eV per atom. SU structures satisfy the MACE relaxation-convergence and energy criteria after structural deduplication; SUN structures additionally remain unmatched to both the Materials Project database and the supplied references structures. (d) Joint t-SNE visualization of 700 supplied reference structures and 220 HiMatGen candidates using a 104-dimensional, species-agnostic geometric fingerprint. Pale circles denote reported structures, and dark stars denote HiMatGen candidates; colors distinguish chemical systems.

## Designing materials under joint functional and chemical constraints

We next applied HiMatGen to three property-directed tasks combining functional targets with chemical or physical constraints (**Figure 3**). Lightweight stiffness design required a bulk modulus $K \geq 80$ GPa at a density $\rho \leq 3.0$ g $cm^{-3}$. Lead-free halide design targeted inorganic compositions within the supported HSE+SOC model domain, with a band gap of 1.5–2.5 eV. Dielectric design required $\kappa \geq 20$ together with $E_g \geq 3.0$ eV in an oxide or oxyfluoride. These tasks call for different chemical strategies: forming stiff networks from lightweight elements, achieving a suitable optical gap without toxic lead, and combining a high dielectric response with electronic insulation.

We similarly assessed each portfolio through structural diversity, SUN structures and target-achieved SUN structures (**Figure 3a–c**). Structural diversity counts unique, in-scope evaluated structures, while target-achieved SUN structures additionally satisfy all joint design requirements. These measures distinguish the breadth of exploration from the number of novel structures meeting the design objective. Final-CIF counts, elemental restrictions and evaluation coverage are reported in **Table S3** and **Figure S6**.

HiMatGen produced 2, 17 and 6 target-achieved SUN structures in lightweight stiffness, lead-free halide and high-κ–wide-gap design, respectively, exceeding both the single Terra agent and Independent SWARM across all three tasks (**Figure 3a–c**). The tool-enabled GPT-5.6 Sol agent produced 0, 7 and 0, respectively. The strongest advantage emerged in lead-free halides, where HiMatGen retained 17 target-achieved SUN structures from 18 unique evaluated structures, compared with two for single Terra, three for Independent SWARM and eight for GPT-6 Astra. In dielectric design, HiMatGen retained six target-achieved SUN structures, close to Astra's seven, whereas Astra performed better in lightweight stiffness, producing seven compared with HiMatGen's two. These results demonstrate that the complete Terra-based HiMatGen workflow outperforms same-model controls across all three tasks and achieves task-dependent advantages over stronger single-agent models.

The halide portfolio also illustrates the distinction between structural and compositional diversity. HiMatGen's 17 target-achieved SUN structures span eight distinct compositions. Because the optical evaluator is composition based, different atomic arrangements at the same composition receive the same band-gap estimate. HiMatGen therefore explored both alternative compositions and distinct structural realizations within those compositions. Results at the broader 0.20 eV per atom threshold are reported in **Figure S7**, and composition-resolved counts are provided in **Figure S8**.

Together, these results identify research organization as an important design variable alongside the underlying model. Hierarchical coordination enabled Terra-based investigators to develop target-achieving portfolios beyond those of the same-model controls and, in lead-free halides, beyond the frontier single-agent baseline. This outcome motivates extending the architecture to stronger models, where improved chemical reasoning could further strengthen proposal development and cross-domain evaluation.

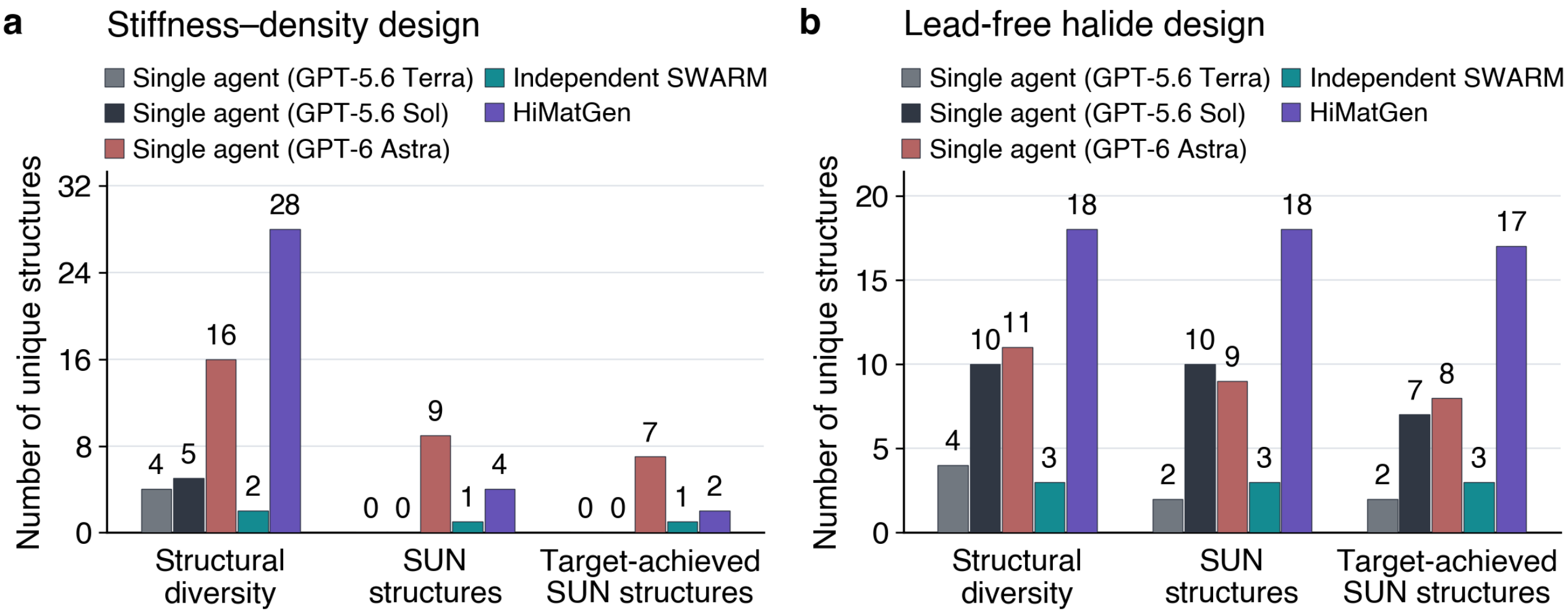

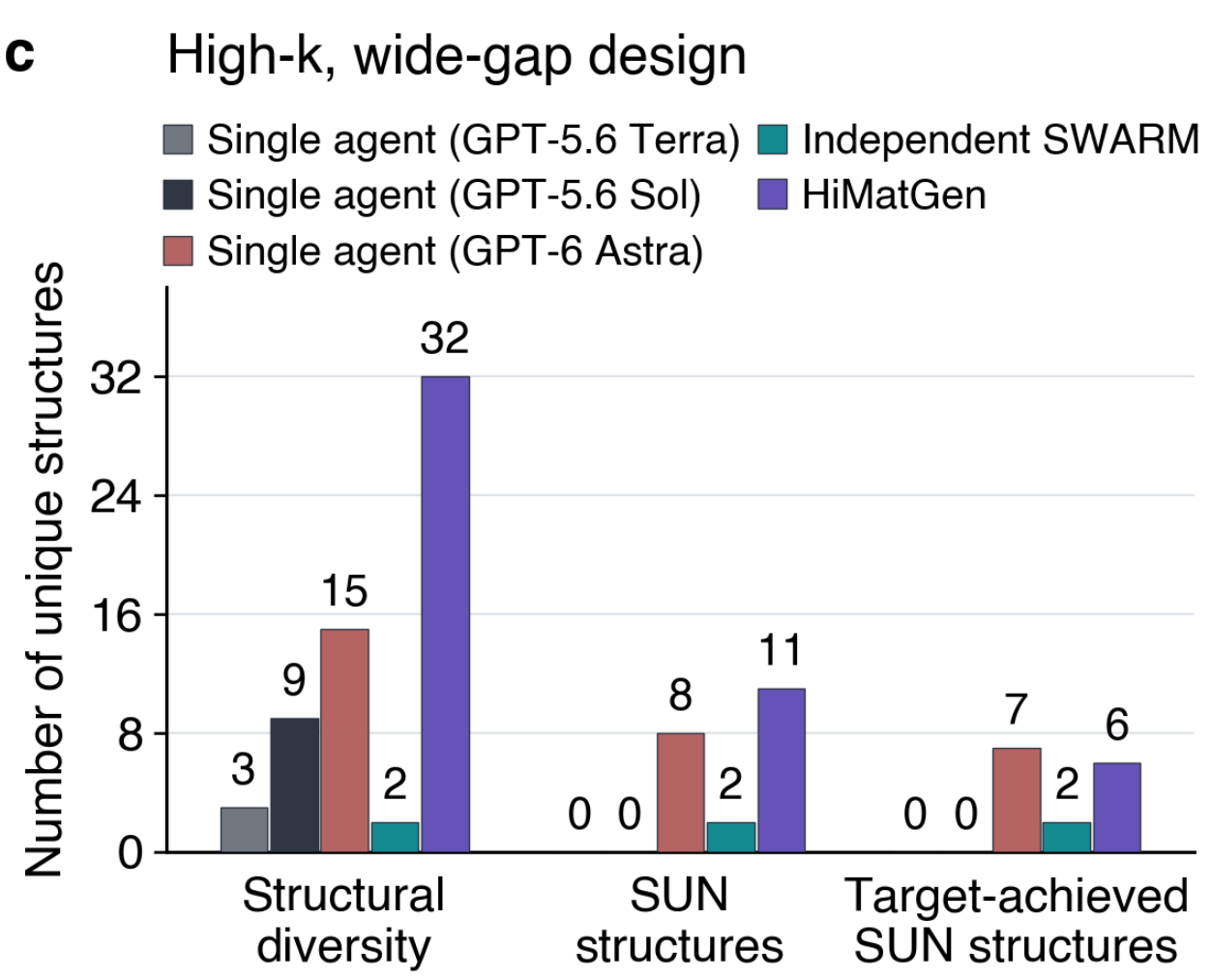

**Figure 3. Tool-grounded materials design under joint functional and chemical constraints.** (a) Lightweight stiffness: bulk modulus K ≥80 GPa and density ρ ≤3.0 g $cm^{-3}$. (b) Lead-free halides: band gap of 1.5–2.5 eV for compositions supported by the HSE+SOC model. (c) High-κ–wide-gap dielectrics: dielectric constant κ ≥20 and band gap Eg ≥3.0 eV. Each panel compares structural diversity, measured as unique in-scope evaluated structures, with SUN and target-achieved SUN counts. Target-achieved SUN structures additionally satisfy all joint design requirements within the same qualifying structural member.

## Turning scientific disagreement into structural alternatives

To examine how hierarchical collective reasoning shaped these design outcomes, we traced two high-κ design trajectories through agent exchanges, tool records and successive crystal structures (**Figure 4**). Both show how questions raised during discussion led to specific structural alternatives and further investigation.

The first trajectory concerns fluorine coordination in $Ba_2ScTiO_5F$ (**Figure 4a**). Following pod-level discussion, the composition representative constructed an initial design with F associated with Sc and asked the structure representative to examine its local environment. Their exchange prompted a Ti–F–Ti alternative, which the structure representative evaluated before sharing the findings with a five-investigator pod. The investigators then proposed a reciprocal Sc–F–Sc control, leading to another construction and evaluation step. Through this upward and downward exchange, an initial oxyfluoride proposal developed into three explicit structural alternatives.

The alternatives addressed distinct questions about anion ordering. The first redesign changed both cation ordering and topology, while the subsequent control exchanged F and O at fixed cell and cation ordering to examine the local fluorine environment more directly. All three met the band-gap and dielectric targets according to the CRISP models. Following MACE relaxation, the common-hull evaluation distinguished their energetics: the initial design and reciprocal control passed the 0.10 eV per atom threshold, whereas the Ti–F–Ti alternative passed at 0.20 eV per atom. Retaining these branches allowed the team's chemical hypotheses to be compared through both their functional properties and energetic qualification.

The second trajectory concerns cation ordering in $HfZrO_4$ (**Figure 4b**). The defect-and-surface representative modified a supplied $HfO_2$ structure to construct a six-site Hf/Zr arrangement, evaluated it and returned the findings to a five-investigator pod. Their replies called for alternative ordering controls, motivating a 12-site checkerboard structure that the representative subsequently constructed and evaluated. Both arrangements met the predicted property targets and passed the common-reference energetic screen. Follow-up review combined synthesis-record retrieval and CRISP-synthesizability with geometric surface analysis. Retrieval identified synthesis analogues without an exact match for the proposed cation order, while a constructed slab exposed undercoordinated Hf sites. These findings prompted requests for additional ordering and surface controls, extending the investigation beyond the initial bulk structures. Additional ordering examples involving $Ba_2LaTaO_6$ and $BaScO_2F$ are presented in **Figure S9**.

These trajectories illustrate how the hierarchy converts scientific discussion into informative structural controls. Feedback from investigators changes which alternatives are constructed, and computational findings shape the questions returned to the pods. This continued exchange allows collective reasoning to influence the evolving crystal structures throughout the design process.

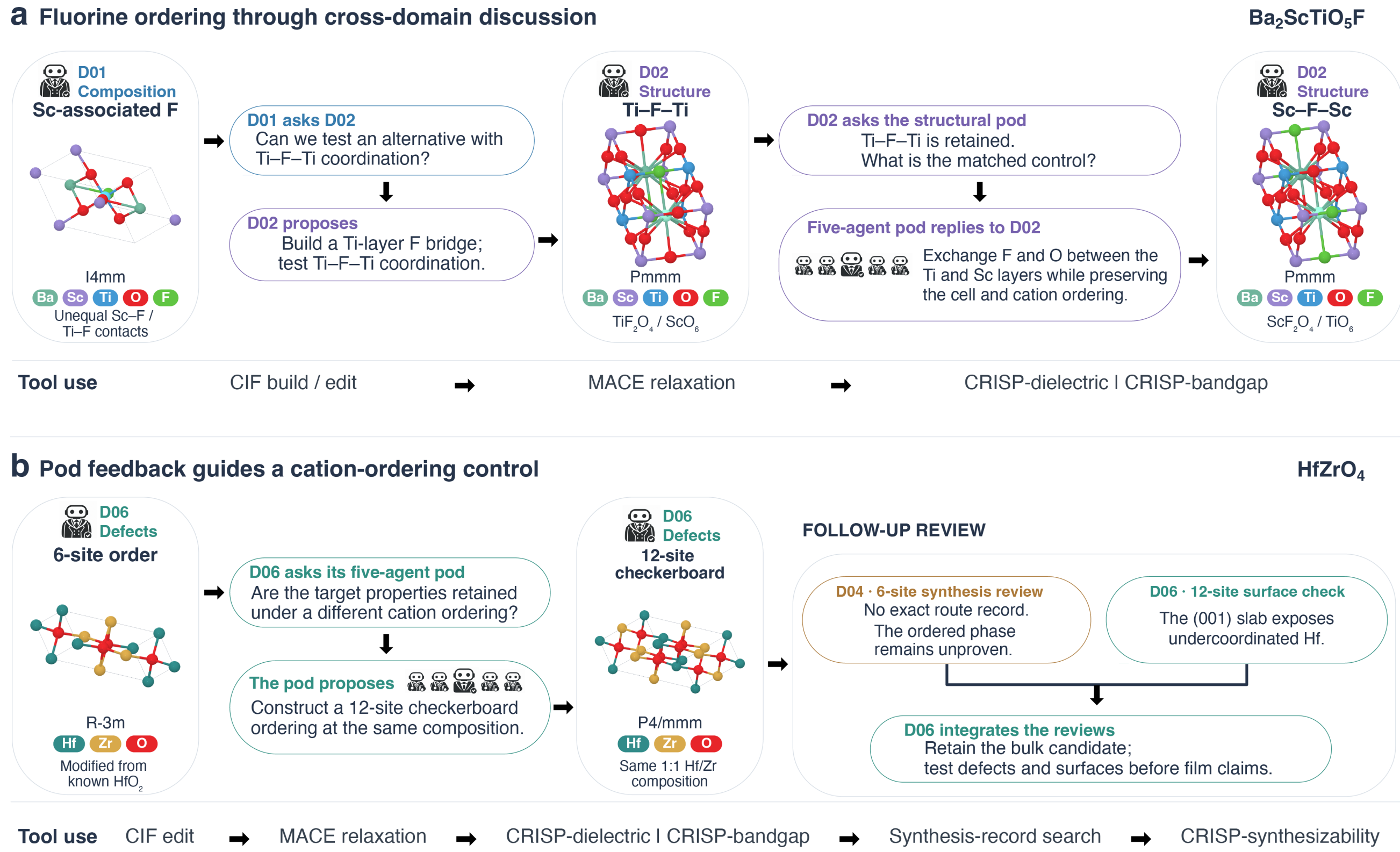

**Figure 4. Cross-domain discussion and pod feedback guide structural revision.** (a) In $Ba_2ScTiO_5F$, an exchange between the composition and structure representatives develops an initial Sc-associated-F design into a Ti–F–Ti alternative. Subsequent feedback from a five-investigator pod prompts a reciprocal Sc–F–Sc control through F/O exchange at fixed cell and cation ordering. (b) In $HfZrO_4$, the defect-and-surface representative constructs a six-site Hf/Zr arrangement from a supplied $HfO_2$ parent. Pod feedback motivates a 12-site checkerboard ordering at the same composition. Follow-up synthesis and surface reviews identify route analogues and undercoordinated Hf sites, prompting requests for further controls. Colored boxes summarize agent exchanges, and separate tool-use rows indicate structure construction, computational evaluation and synthesis assessment. Exchanges are paraphrased for clarity; verbatim excerpts and CIF lineage are provided in **Supplementary Note 4**.

## Comparing agent-guided and generative approaches to materials design

We compared the property space explored by HiMatGen with that of MatterGen[4] and Chemeleon 2[3], each generating 1,000 structures per task across ten seeds of 100 structures. This substantially larger sampling budget gave the generative models ample opportunity to explore the target regions, allowing us to examine their coverage of the joint design requirements alongside HiMatGen's smaller, research-driven portfolios rather than rank the methods at a matched output count. MatterGen used supported bulk-modulus or band-gap conditioning, whereas Chemeleon 2 used the public unconditional RL-DNG checkpoint. All outputs were processed using the common evaluation and structural-matching protocol. **Figure 5a–c** maps SUN structures with available property evaluations. Complete generation counts and evaluation coverage are reported in **Figures S6** and **S10–S13**.

Both generative models populated a broad stiffness–density landscape (**Figure 5a**). Following SUN screening, the plot contains 774 MatterGen and 899 Chemeleon 2 structural groups, compared with 4 from HiMatGen. Despite this smaller SUN portfolio, HiMatGen's candidates were concentrated near the joint design target: 2 satisfied both requirements, while the remaining 2 lay close to the target boundary. The target-achieving structures, $B_{12}NO$ and $B_6N$ (**Figure 5d,g**), reflect a chemically directed strategy combining light elements with connected bonding networks. This concentration illustrates the capacity of agent-guided exploration to develop bespoke candidates around coupled design requirements, complementing the broad structural coverage provided by generative models.

Lead-free-halide design imposed more specific compositional requirements (**Figure 5b**). Under these combined chemical and optical constraints, MatterGen yielded only 1 target-achieved SUN structure, $KGeCl_3$, while Chemeleon 2 yielded none with an applicable band-gap evaluation. HiMatGen retained 17 target-achieved SUN structures, demonstrating its ability to focus exploration on the requested chemistry and develop multiple structural alternatives within the target window. Representative candidates $CsSnIBr_2$ and $CsGeIBr_2$ illustrate how mixed-halide composition and B-site substitution supported this targeted design strategy (**Figure 5e,h**). In high-κ–wide-gap design, HiMatGen produced 6 target-achieved SUN structures, compared with 8 for MatterGen and none for Chemeleon 2 (**Figure 5c**). HiMatGen's $Ba_2HfZrO_6$ and $Sr_2HfZrO_6$ candidates further illustrate the development of related cation designs satisfying the joint dielectric and band-gap requirements (**Figure 5f,i**).

These comparisons highlight complementary strengths. Generative models supplied broad populations of energetically qualified, reference-unmatched structures, while HiMatGen developed chemically targeted alternatives through tool-guided investigation and structural revision. Its design records connect candidate structures to the hypotheses, evaluations and decisions underlying their development. The supplementary absorber task further illustrates MatterGen's strength in single-property conditioning (**Figure S11**). Together, these results demonstrate the value of hierarchical collective reasoning for bespoke materials design, translating joint functional and chemical requirements into targeted structural alternatives.

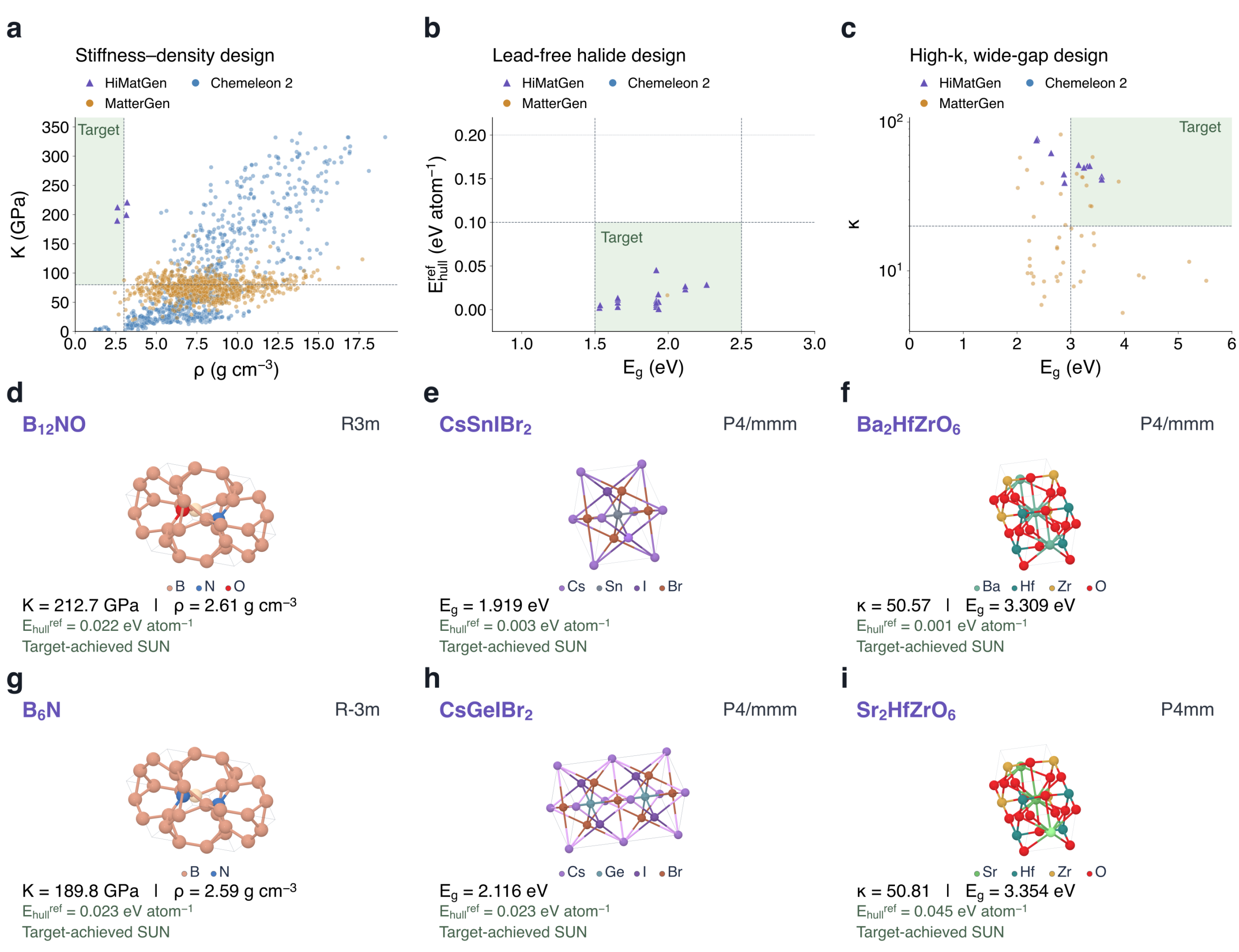

**Figure 5. Property-space exploration and representative target-achieving designs.** HiMatGen is compared with MatterGen and Chemeleon 2, each generating 1,000 structures per task across ten seeds of 100 structures. (a) Lightweight stiffness design: 4 HiMatGen, 774 MatterGen and 899 Chemeleon 2 SUN structural groups. (b) Lead-free halide design: 17 HiMatGen and 1 MatterGen target-achieved SUN structural group; no Chemeleon 2 group satisfies the inclusion criteria. (c) High-κ–wide-gap design: 11 HiMatGen and 45 MatterGen SUN structural groups; no Chemeleon 2 group satisfies the inclusion criteria. Shaded regions indicate the joint design targets. (d–i) Representative HiMatGen designs for three tasks.

## Scaling collective reasoning through local communication

Expanding the investigator population creates more opportunities for exploration but also increases the volume of information exchanged. We examined this communication burden in a paired Al–Cu–Mg experiment initialized from the same 100 independent reports. Both conditions completed two debate rounds, producing 200 investigator responses. HiMatGen shared reports within five-investigator pods, whereas full debate supplied each investigator with the other 99 reports. Each condition then proceeded to tool-grounded investigation and integration.

Local discussion required 800 directed peer-report deliveries, compared with 19,800 for full debate—a 24.75-fold difference (**Figure 6a**). Debate-stage input-plus-output usage was 2.52 million tokens for HiMatGen and 55.44 million for full debate, corresponding to a 21.98-fold reduction. Most of this difference arose from repeated input: the mean input per response was 10,645 tokens for local discussion and 275,011 for full debate. Output volumes were much closer, totaling 393,149 and 441,281 tokens, respectively. Local communication therefore substantially reduced repeated exposure to existing reports while preserving a similar volume of newly generated discussion.

HiMatGen also retained a larger structural portfolio in the paired experiment (**Figure 6b**). Applying the same audit to all successfully generated structures, including intermediate artifacts, yielded 18 SU structures for HiMatGen and 5 for full debate at the 0.10 eV per atom threshold. Because downstream investigation used ten tool-enabled representatives in HiMatGen and one tool-enabled integrator in full debate, this quality comparison reflects the complete research organizations rather than communication topology alone. Across the complete workflows, token usage was 43.33 million for HiMatGen and 82.57 million for full debate, a 1.91-fold difference after including initial generation, subsequent investigation and final integration (**Figures S14** and **S15**).

The communication advantage follows from limiting the number of recipients for each report. With N investigators, r debate rounds and pod size p, local sharing requires $rN(p-1)$ directed deliveries, compared with $rN(N-1)$ for all-to-all debate. At fixed pod size and round count, local report sharing therefore grows linearly rather than quadratically with the investigator population. Representative-level investigation and feedback contribute additional communication according to the research pursued. Across the six chemical systems, mean SUN counts increased from 3.33 at N = 10 to 3.67, 6.33, 8.00 and 8.33 at N = 20, 50, 100 and 200, respectively (**Figure 6c**). These results show substantial expansion up to 100 investigators, followed by a smaller additional gain at 200 under the evaluated settings.

A practical limitation of HiMatGen is its longer end-to-end runtime. Across the six chemical systems, HiMatGen averaged 93.22 min per trajectory, compared with 22.01 min for Independent SWARM, 5.29 min for single Terra, 68.81 min for single Sol and 71.96 min for single Astra (**Figure S16**). These recorded times include interruptions and recovery waits alongside model and tool execution. Sustained investigation across multiple research stages therefore introduces a time–breadth trade-off: HiMatGen develops a broader structural portfolio but takes longer to complete the design process. Local communication reduces the overhead of supporting this extended investigation, making hierarchical coordination a practical approach when exploration breadth is prioritized over rapid turnaround.

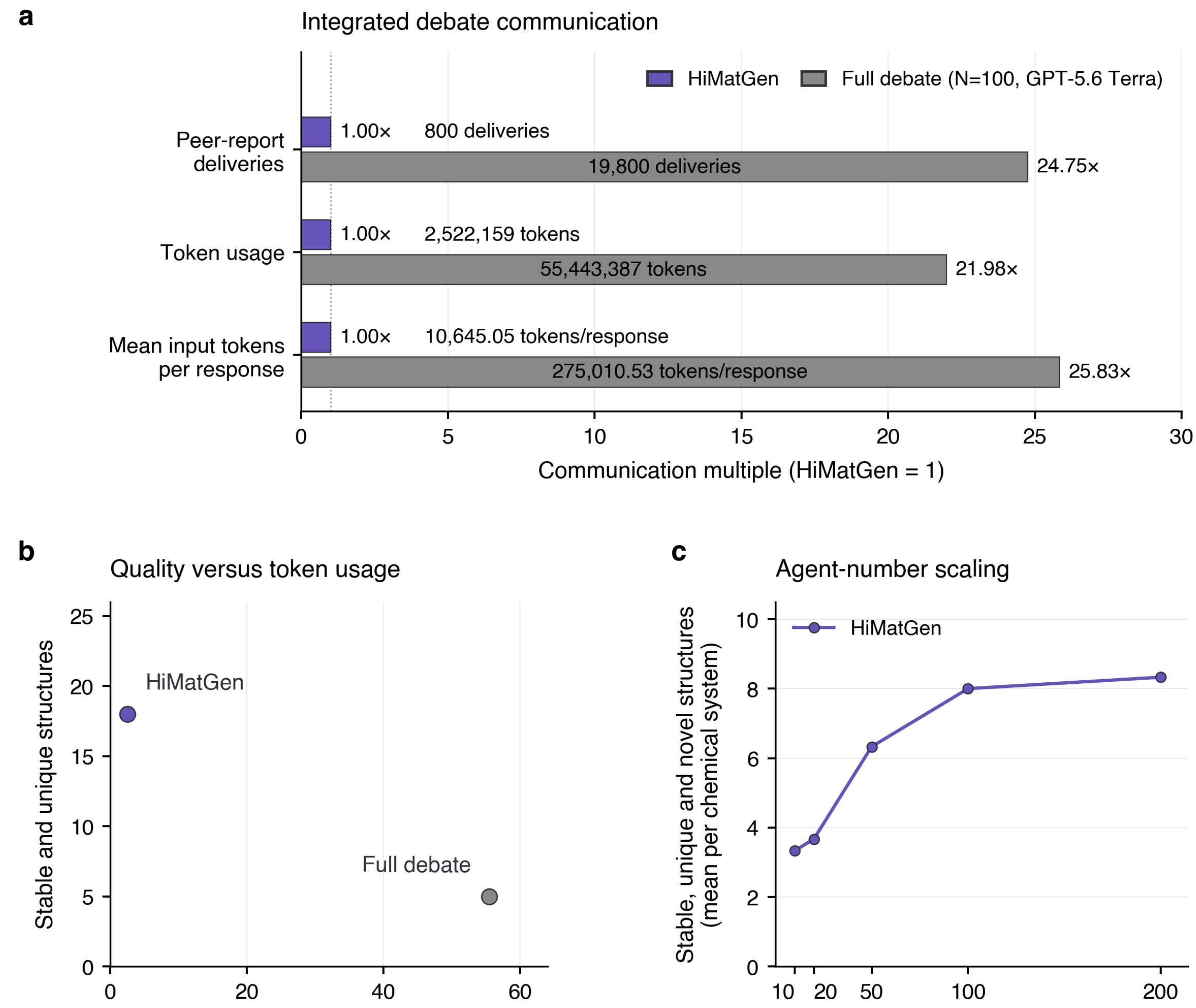

**Figure 6. Communication efficiency and scaling of collective reasoning.** (a) Directed peer-report deliveries, debate-stage input-plus-output tokens and mean input per debate response, each normalized to HiMatGen = 1, with absolute values annotated. (b) Stable and unique (SU) structures versus debate-stage input-plus-output tokens. Both methods are evaluated using the same all-generated-structure audit, including successfully generated intermediate structures, at an energy-above-hull threshold of 0.10 eV per atom. (c) Mean stable, unique and novel (SUN) structure counts across six chemical systems at $N = 10$, 20, 50, 100 and 200 investigators.

# CONCLUSIONS

HiMatGen establishes hierarchical collective reasoning as a framework for scaling LLM-based materials design. Across six chemical systems, it retained an average of 8 SUN structures per system, exceeding the single-agent and independent-generation baselines, and extended this capability to design tasks combining functional targets with chemical constraints. Agent exchanges showed how competing interpretations of fluorine coordination and cation ordering led to explicit structural alternatives and controls. Local discussion reduced debate-stage token usage by approximately 22-fold relative to full debate, supporting broad participation without repeatedly broadcasting every argument to every investigator.

The central contribution is an organization in which scientific hypotheses remain open to development throughout the research process. Domain representatives connect independent proposals to computational evidence, seek complementary expertise and return unresolved questions to the investigators. This exchange turns disagreement into testable alternatives and allows findings from one branch to redirect another. This iterative process supports design goals expressed through open-ended natural-language instructions, combining functional targets, chemical restrictions and qualitative preferences beyond a predefined set of generative-model conditioning variables. The emergence of highly capable models such as GPT-6 Astra makes the organization of collective reasoning increasingly consequential: stronger individual reasoning expands what an agent can investigate, while coordination determines how multiple investigations inform one another. Our results show that HiMatGen can achieve competitive, and in some tasks superior, design outcomes relative to a stronger single agent. Together, these findings position research organization as a complementary axis of progress alongside model capability and motivate extending hierarchical collective reasoning to more powerful models.

Several directions remain open. The present ten-domain structure is a practical, expert-informed organization rather than an established optimum; its composition and coordination may need to adapt to different materials problems. Human expertise remains important for defining meaningful objectives, selecting appropriate tools and judging which hypotheses warrant further investigation. The proposed structures also require higher-fidelity calculations and experimental validation to establish their stability, synthesizability and functional performance. These next steps offer an opportunity to extend the same feedback principle beyond computational screening, allowing experimental observations to revise hypotheses and guide subsequent design. Applying the architecture to stronger models could further improve proposal development and cross-domain review. More broadly, HiMatGen provides a foundation for materials research in which distributed knowledge supports the generation, testing and refinement of chemical hypotheses, yielding both candidate structures and a traceable basis for deciding what to investigate next.